\documentclass[envcountsame]{llncs}

\usepackage[dvipsnames]{xcolor}  
\usepackage[%
    framemethod=tikz,
    skipbelow=\topskip,
    skipabove=\topskip
]{mdframed}
\usepackage[utf8]{inputenc}
\usepackage{amsmath}
\usepackage{amssymb}
\usepackage{mathtools}
\usepackage{hyperref}
\usepackage[inline]{enumitem}
\usepackage{array}
\usepackage{balance}
\usepackage{algorithm}
\usepackage{xfrac}
\usepackage[noend]{algpseudocode}
\usepackage{xspace}
\usepackage{subcaption}
\usepackage{listings}
\usepackage{xparse}
\usepackage{etoolbox}
\usepackage{smartdiagram}
\usepackage{floatflt}
\usepackage{mdframed,framed}
\usepackage[font=small]{caption}

\let\doendproof\endproof
\renewcommand\endproof{\qed\doendproof}

\let\set\mathbb

\newcommand{\ass}{\mathrel{:=}}     
\newcommand{\reserved}[1]{\textbf{#1}} 
\newcommand{\DO}{\reserved{do}}
\newcommand{\OD}{\reserved{end~while}}
\newcommand{\WHILE}{\reserved{while}}
\newcommand{\IF}{\reserved{if}}
\newcommand{\FI}{\reserved{end~if}}
\newcommand{\THEN}{\reserved{then}}

\newcommand{\ELSE}{\reserved{else}}

\newcommand{\package}[1]{{\small\texttt{#1}}}
\newcommand{\Aligator}{\package{Aligator}\xspace}
\newcommand{\AligatorJL}{\package{Aligator.jl}\xspace}
\newcommand{\Julia}{Julia\xspace}
\newcommand{\Sage}{SageMath\xspace}

\newcommand{\cpp}{C\texttt{++}\xspace}

\definecolor{bggray}{gray}{0.97}
\definecolor{repl}{RGB}{2, 114, 4}

\mdfsetup{%
    leftmargin=0pt,
    rightmargin=0pt,
    backgroundcolor=bggray,
    linewidth=0,
    roundcorner=2
}

\BeforeBeginEnvironment{lstlisting}{\vskip-4pt\begin{mdframed}\vspace{-0.7em}}
\AfterEndEnvironment{lstlisting}{\vspace{-0.7em}\end{mdframed}\vskip-7pt}

\lstdefinelanguage{julia}
{
  keywordsprefix=\&,
  morekeywords={
    exit,whos,edit,load,is,isa,isequal,typeof,tuple,ntuple,uid,hash,finalizer,convert,promote,
    subtype,typemin,typemax,realmin,realmax,sizeof,eps,promote_type,method_exists,applicable,
    invoke,dlopen,dlsym,system,error,throw,assert,new,Inf,Nan,pi,im,begin,while,for,in,return,
    break,continue,macro,quote,let,if,elseif,else,try,catch,end,bitstype,ccall,do,using,module,
    import,export,importall,baremodule,immutable,local,global,const,Bool,Int,Int8,Int16,Int32,
    Int64,Uint,Uint8,Uint16,Uint32,Uint64,Float32,Float64,Complex64,Complex128,Any,Nothing,None,
    function,type,typealias,abstract
  },
  sensitive=true,
  morecomment=[l]{\#},
  morestring=[b]',
  morestring=[b]" 
}

\let\oldsubsection\subsection
\renewcommand\subsection[2][]{%
\vspace{-4pt}%
\oldsubsection[#1]{#2}%
}

\algrenewcommand\algorithmicrequire{\textbf{Input:}}
\algrenewcommand\algorithmicensure{\textbf{Output:}}

\algnewcommand\algorithmicand{\textbf{\textsf{AND}}}
\algrenewcommand\algorithmicwhile{\textbf{\textsf{WHILE}}}
\algrenewcommand\algorithmicfor{\textbf{\textsf{FOR}}}
\algrenewcommand\algorithmicdo{\textbf{\textsf{DO}}}
\algrenewcommand\algorithmicreturn{\textbf{\textsf{RETURN}}}

\title{Aligator.jl -- A Julia Package for Loop Invariant Generation \thanks{All authors are supported by the ERC Starting Grant 2014 SYMCAR
    639270. We also acknowledge funding from the Wallenberg Academy
    Fellowship 2014 TheProSE, the Swedish VR grant GenPro D0497701, and the
    Austrian FWF research projects RiSE S11409-N23 and
    W1255-N23. Maximilian Jaroschek is also supported by the FWF
    project Y464-N18.}}

\author{Andreas Humenberger\inst{1} \and Maximilian
  Jaroschek\inst{1,2}\and Laura Kov\'acs\inst{1,3}}
\institute{
  TU Wien \\\email{ahumenbe@forsyte.at}
  \and
  JKU Linz\
\and 
Chalmers\vspace{-1em}}

\begin{document}

\urlstyle{tt}
\maketitle
\begin{abstract}
  We describe the \AligatorJL software package for automatically generating all
  polynomial invariants of the rich class of extended P-solvable loops with
  nested conditionals. \AligatorJL is written in the programming language \Julia
  and is open-source. \AligatorJL transforms program loops into a system of
  algebraic recurrences and implements techniques from symbolic computation to
  solve recurrences, derive closed form solutions of loop variables and infer
  the ideal of polynomial invariants by variable elimination based on Gr\"obner
  basis computation. 
\end{abstract}


\section{Introduction}

In~\cite{DBLP:conf/vmcai/HumenbergerJK18}
we described an automated approach for generating loop invariants as a
conjunction of polynomial equalities for a family of loops, called extended
P-solvable loops. For doing so, we abstract loops to  
a system of algebraic recurrences over the loop counter and
program variables and compute polynomial equalities among loop
variables from the closed form solutions of the recurrences. 
%
%

\vspace{0.7ex}
\noindent\textbf{Why Julia?}
Our work was previously implemented in the \Aligator software
package~\cite{aligator}, within the Mathematica system~\cite{Mathematica}.
While Mathematica provides high-speed implementations of symbolic computation
techniques, it is a proprietary software which is an obstacle for using
\Aligator{} in applications of invariant generation. The fact that Mathematica provides no possibility to parse and modify
program code was also a reason to move to another environment. To make \Aligator
better suited for program analysis, we decided to redesign \Aligator in the
\Julia programming language~\cite{Julia}.
\Julia provides a simple and efficient interface for calling C/\cpp and Python
code. This allows us to resort to already existing computer algebra libraries,
such as Singular~\cite{singular} and SymPy~\cite{sympy}. \Julia also provides a
built-in package manager that eases the use of other packages and enables others
to use Julia packages, including our \AligatorJL tool.
Before committing to \Julia, we also considered the computer algebra system
\Sage~\cite{Sage} and an implementation directly in C/\cpp as options for
redesigning \Aligator. The former hosts its own Python version which makes the
installation of other Python packages (e.g.~for parsing source code) tedious and
error-prone. While C/\cpp is very efficient and provides a large ecosystem on
existing libraries, developing C/\cpp projects requires more effort than \Julia
packages. We therefore believe that \Julia provides the perfect mix between
efficiency, extensibility and convenience in terms of programming and symbolic
computations.

%

\vspace{0.7ex}
\noindent\textbf{\AligatorJL.} This paper overviews \AligatorJL and details its main components. The code of
\AligatorJL is available open-source at:
\begin{center}
  \url{https://github.com/ahumenberger/Aligator.jl}.
\end{center} 
All together, \AligatorJL consists of about 1250 lines of Julia code. We
evaluated \AligatorJL on challenging benchmarks on invariant generation. Our
experimental results are available at the above mentioned link and demonstrate
the efficiency of \AligatorJL. 
\paragraph{\bf Contributions.} Our new tool \AligatorJL{} significantly extends and improves the
existing software package \Aligator{} as follows: 
\begin{itemize}
\item Unlike \Aligator{}, \AligatorJL is open-source and easy to
  integrate into other software packages. 
\item \AligatorJL{} implements symbolic computation techniques
  directly in Julia for extracting
  and solving recurrences and
  generates polynomial dependencies among exponential sequences.
\item  Contrarily to \Aligator{}, \AligatorJL{} handles not
  only linear recurrences with constant coefficients, called C-finite
  recurrences. Rather, \AligatorJL also supports hypergeometric sequences and
  sums and term-wise products of C-finite and hypergeometric recurrences~\cite{DBLP:conf/vmcai/HumenbergerJK18}.
\item  \AligatorJL is complete. That is, a 
  finite basis of the polynomial invariant ideal is always
  computed.
\end{itemize}

\section{Background and Notation}\label{sec:background}

\AligatorJL computes polynomial invariants of so-called extended P-solvable
loops~\cite{DBLP:conf/vmcai/HumenbergerJK18}. Loop guards and test conditions are ignored in such loops and denoted by $\ldots$ or
\lstinline{true}, yielding non-deterministic loops with sequencing and conditionals. 
Program variables $V=\{v_1,\ldots,v_m\}$ of extended
P-solvable loops have numeric values, abstracted to be rational numbers. The
assignments of extended P-solvable loops are of the form
$v_i\ass \sum_{j=0}^m c_iv_j+c_{m+1}$ with constants $c_0,\dots,c_{m+1}$, or
$v_i\ass r(n)v_i$, where $r(n)$ is a rational function in the loop counter $n$.
We give an example of an
extended P-solvable loops in
Figure~\ref{fig:extPsolvable}. 
\begin{floatingfigure}[r]{53mm}
\vspace*{-1.4em}
\begin{mdframed}[
  userdefinedwidth=53mm, 
  innertopmargin=13pt, 
  innerbottommargin=12pt, 
  linewidth=.4px, 
  rightmargin=0pt, 
  backgroundcolor=white, 
  roundcorner=0pt
]
\begin{tabular}{l}
\WHILE\ $\dots$\ \DO\\
    \quad\IF\ $\dots$\ \THEN \\
    \qquad $r\ass  r - v$; $v\ass  v + 2$\\ 
    \quad\ELSE\\
    \qquad $r\ass  r + u$; $u\ass  u + 2$\\ 
    \quad\FI\\
    \OD
\end{tabular}
\caption{{An extended P-solvable loop.}}
\label{fig:extPsolvable}
\end{mdframed}
\vspace*{-1.4em}
\end{floatingfigure}

In correspondence to $V\kern-2pt,$ the initial values of the variables are given 
by the set $V_0:=\{v_1(0), \ldots,v_m(0)\}$; that is, $v_i(0)$ is the initial
value of $v_i$. In what follows, we consider $V$ and $V_0$ fixed and state all
definitions relative to them.  Given an extended P-solvable loop as input,
\AligatorJL generates all its polynomial equality invariants. By a
polynomial equality invariant, in the sequel simply polynomial invariant, we
mean the equality:
\begin{equation}
  \label{eq:polyinv}
  p(v_1,\dots,v_m,v_1(0),\dots,v_m(0))=0,
\end{equation}
where $p$ is a polynomial in $V\cup V_0$ with rational number coefficients.
In what follows, we also refer to the polynomial $p$ in~\eqref{eq:polyinv} as a
polynomial invariant. For $n\in\set N\setminus\{0\}$ and a loop variable $v_i$,
we write $v_i(n)$ to denote the value of $v_i$ after the $n$th loop
iteration. As~\eqref{eq:polyinv} is a loop invariant, we have:
\[p(v_1(n),\dots,v_m(n),v_1(0),\dots,v_m(0))=0\text{ for  } n>0.\]

As shown in~\cite{kapur,DBLP:conf/vmcai/HumenbergerJK18}, the set of polynomial
invariants in $V\kern-2pt$, w.r.t.~the initial values $V_0$, forms a polynomial
ideal, called the polynomial invariant ideal. Given an extended P-solvable loop,
\AligatorJL computes \emph{all} its polynomial invariants as it computes a
basis of the polynomial invariant ideal, a finite set of polynomials
$\{b_1,\dots,b_k\}$. Any polynomial invariant can be written as a linear
combination $p_1b_1+\dots+p_kb_k$ for some polynomials
$p_1,\dots,p_k$. 

\section{System Description of \texttt{Aligator.jl}}

Inputs to \AligatorJL are extended P-solvable loops and are fed to
\AligatorJL 
as  \lstinline{String} in the \Julia syntax. We illustrate the use of
\AligatorJL on our example from Figure~\ref{fig:extPsolvable}: \label{page:install}
%
%
\begin{example}\label{ex:loopstr}
Figure~\ref{fig:extPsolvable} is specified as a Julia string as
follows: 
\begin{lstlisting}
@\repl@ loopstr = """
            while true
                if true
                    r = r - v; v = v + 2
                else
                    r = r + u;  u = u + 2
                end
            end
        """
\end{lstlisting}
\vspace{-0.5em}
\end{example}
%
%
Polynomial loop invariants are inferred using
\AligatorJL by calling the function
\lstinline{aligator(str::String)} with a string input containing the loop as its
argument.
\begin{lstlisting}
@\repl@ aligator(loopstr)
Singular Ideal over Singular Polynomial Ring (QQ),(r_0,v_0,u_0,r,v,u)
  with generators (v_0^2-u_0^2-v^2+u^2+4*r_0-2*v_0+2*u_0-4*r+2*v-2*u)
\end{lstlisting}
%
%
%
The result of \AligatorJL is a Gr\"obner basis of the polynomial invariant
ideal.
It is  represented as an object of type
\lstinline{Singular.sideal} that is defined in the Singular package.
For Figure~\ref{fig:extPsolvable}, \AligatorJL reports that the polynomial invariant ideal is
generated by the polynomial invariant $\{v_0^2-u_0^2-v^2+u^2+4r_0-2v_0+2 u_0 - 4 r+ 2 v - 2 u =0\}$ in variables
$r_0, v_0, u_0, r,v,u$, where  $r_0,v_0,u_0$ denote respectively the
initial values of $r,v,u$. \label{page:aligatorOut}



We now overview the main parts of \AligatorJL: (i)
extraction of
recurrence equations, (ii) recurrence solving and (iii) computing the
polynomial invariant ideal.
%

\paragraph{\bf Extraction of Recurrences.} Given an extended
P-solvable loop as a \Julia string,
\AligatorJL creates the abstract syntax tree of this loop. This tree is then traversed in order
to extract loop paths (in case of a multi-path loop) and the
corresponding loop assignments. The resulting structure is
then flattened in order to get a loop with just one layer of nested
loops. 
Within \AligatorJL this is obtained via the method
\lstinline{extract_loop(str::String)}. As a result, 
the extracted recurrences are represented in \Aligator by an object of
type \lstinline{Aligator.MultiLoop}, in case the input is a multi-path
loop; otherwise, the returned object is of type
\lstinline{Aligator.SingleLoop}.



\begin{example}\label{ex:loop}
Using  Example~\ref{ex:loopstr},
\AligatorJL derives the loop and its corresponding systems of recurrences:
\begin{lstlisting}
@\repl@ loop = extract_loop(loopstr)
2-element Aligator.MultiLoop:
 [r(n1+1) = r(n1) - v(n1), v(n1+1) = v(n1) + 2, u(n1+1) = u(n1)]
 [r(n2+1) = r(n2) + u(n2), u(n2+1) = u(n2) + 2, v(n2+1) = v(n2)]
\end{lstlisting}
As loop paths are translated into single-path loops,  \AligatorJL introduces a loop counter for each
path and computes the recurrence equations of the loop variables $r,v,u$ 
with respect to the loop counters $n_1$ and $n_2$.
\end{example}

\paragraph{\bf Recurrence Solving.}
For each single-path loop, its system of recurrences is solved. \AligatorJL performs various simplifications on the
extracted recurrences, for example by eliminating cyclic dependencies
introduced by auxiliary variables and uncoupling mutually dependent
recurrences. The resulting, simplified recurrences represent sums and term-wise
products of C-finite or hypergeometric sequences. 
\AligatorJL computes closed forms solutions of 
such recurrences  by calling the method \lstinline{closed_forms} and using the symbolic manipulation
capabilities of \package{SymPy.jl}: 

\begin{example}\label{ex:CForms}
For Example~\ref{ex:loop}, we get the
following systems of closed forms:
\begin{lstlisting}
@\repl@ cforms = closed_forms(loop)
2-element Array{Aligator.ClosedFormSystem,1}:
 [v(n1) = 2*n1+v(0), u(n1) = u(0), r(n1) = -n1^2-n1*(v(0)-1)+r(0)]
 [u(n2) = 2*n2+u(0), v(n2) = v(0), r(n2) = n2^2+n2*(u(0)-1)+r(0)] 
\end{lstlisting}
The returned value is an array of type \lstinline{Aligator.ClosedFormSystem}.
\end{example}

\paragraph{\bf Invariant Ideal Computation.}
Using the closed form solutions for (each) single-path loop, \AligatorJL
next derives a basis of the  polynomial invariant ideal of the
(multi-path) extended P-solvable loop. 
To this end, \AligatorJL{} uses the \package{Singular.jl} package  for
Gr\"obner basis computations in order to eliminate variables in the
loop counter(s) from the system of closed forms. For multi-path loops,
\AligatorJL relies on iterative Gr\"obner basis computations until
a fixed point is derived representing a Gr\"obner basis of the
polynomial invariant ideal -- see~\cite{DBLP:conf/vmcai/HumenbergerJK18} for theoretical details. 

Computing
polynomial invariants within \AligatorJL is performed by the function
  \lstinline|invariants(cforms::Array{ClosedFormSystem,1})|. 
The result  is an object of type
\lstinline{Singular.sideal} and represents a Gr\"obner basis of the
polynomial invariant ideal in the loop variables. 

\begin{example}\label{ex:invIdeal}
For  Example~\ref{ex:CForms}, \AligatorJL
generates the following Gr\"obner
basis, as already described on page~\pageref{page:aligatorOut}: 
\begin{lstlisting}
@\repl@ ideal = invariants(cforms)
Singular Ideal over Singular Polynomial Ring (QQ),(r_0,v_0,u_0,r,v,u)
  with generators (v_0^2-u_0^2-v^2+u^2+4*r_0-2*v_0+2*u_0-4*r+2*v-2*u)
\end{lstlisting}
\vspace{-1em}
\end{example}

%
%


\section{Experimental Evaluation}

Our approach to invariant generation was shown to outperform
state-of-the-art tools on invariant generation for multi-path loops
with polynomial arithmetic~\cite{DBLP:conf/vmcai/HumenbergerJK18}. 
In this section we focus on the performance of our new implementation
in \AligatorJL and compare results to \Aligator~\cite{aligator}. 
In our experiments, we used benchmarks
from~\cite{DBLP:conf/vmcai/HumenbergerJK18}. Our experiments were performed on a
machine with a 2.9 GHz Intel Core i5 and 16 GB LPDDR3 RAM. When using
\AligatorJL, the invariant ideal computed by \AligatorJL was non-empty for each
example; that is, for each example we were able to find non-trivial invariants.

Tables~(\ref{tab:singlepath}) and~(\ref{tab:multipath}) show the results for a
set of single- and multi-path loops respectively. In both tables the first column
shows the name of the instance, whereas columns two and three depict the running
times (in seconds) of \Aligator and \AligatorJL, respectively.

\newcolumntype{R}{>{$}r<{$}}

\begin{table}
  \small
  \centering
  \caption{Experimental evaluation of \AligatorJL.}
\vspace{-0.5em}  
  \label{tab:benchmarks}
  \def\arraystretch{1.08}
  \begin{subtable}[t]{.45\textwidth}
  \caption{}
  \label{tab:singlepath}
  \begin{tabular}{|l|R|R|}
    \hline
    \textit{Single-path~} & ~\Aligator & ~\AligatorJL \\
    \hline
    \texttt{cohencu}  & 0.072   & 2.879  \\\hline
    \texttt{freire1}  & 0.016   & 1.159  \\\hline
    \texttt{freire2}  & 0.062   & 2.540  \\\hline
    \texttt{petter1}  & 0.015   & 0.876  \\\hline
    \texttt{petter2}  & 0.026   & 1.500  \\\hline
    \texttt{petter3}  & 0.035   & 2.080  \\\hline
    \texttt{petter4}  & 0.042   & 3.620  \\\hline
  \end{tabular}
  \end{subtable}
  \quad
  \begin{subtable}[t]{.45\textwidth}    
  \caption{}
  \label{tab:multipath}  
  \begin{tabular}{|l|R|R|}
    \hline
    \textit{Multi-path~} &  ~\Aligator & ~\AligatorJL  \\\hline
    \texttt{divbin}     &    0.134     & 1.760\\\hline
    \texttt{euclidex}   &    0.433     & 3.272\\\hline
    \texttt{fermat}     &    0.045     & 2.159\\\hline
    \texttt{knuth}      &    ~55.791   & 12.661 \\\hline
    \texttt{lcm}        &    0.051     & 2.089\\\hline
    \texttt{mannadiv}   &    0.022     & 1.251\\\hline
    \texttt{wensley}    &    0.124     & 1.969\\\hline
  \end{tabular}
  \end{subtable}
\end{table}

By design, \AligatorJL is at least as strong as \Aligator concerning the quality
of the output.  When it comes to efficiency though, we note that \AligatorJL is
slower than \Aligator. We expected this result as \Aligator uses the highly
optimized algorithms of Mathematica.  When taking a closer look at how much time
is spent in the different parts of \AligatorJL, we observed that the most time
in \AligatorJL is consumed by symbolic manipulations. Experiments indicate that
we can improve the performance of \AligatorJL considerably by using the Julia
package \package{SymEngine.jl} instead of \package{SymPy.jl}. We believe that
our initial experiments with \AligatorJL are promising and demonstrate the use
of our efforts in making our invariant generation open-source.


\section{Conclusion}

We introduced the new package \AligatorJL for loop invariant generation in the
programming language \Julia. 
Our \AligatorJL tool is an open-source software package for
invariant generation using symbolic computation and can easily be
integrated with other libraries and tools.


\balance
\bibliographystyle{splncs04}
\bibliography{references}

\begin{thebibliography}{1}
\providecommand{\url}[1]{\texttt{#1}}
\providecommand{\urlprefix}{URL }
\providecommand{\doi}[1]{https://doi.org/#1}

\bibitem{singular}
Decker, W., Greuel, G.M., Pfister, G., Sch\"onemann, H.: {\sc Singular} {4-1-0}
  --- {A} computer algebra system for polynomial computations.
  \url{http://www.singular.uni-kl.de} (2016)

\bibitem{DBLP:conf/vmcai/HumenbergerJK18}
Humenberger, A., Jaroschek, M., Kov{\'{a}}cs, L.: {Invariant Generation for
  Multi-Path Loops with Polynomial Assignments}. In: {VMCAI}. Lecture Notes in
  Computer Science, vol. 10747, pp. 226--246. Springer (2018).
  \doi{10.1007/978-3-319-73721-8\_11}

\bibitem{Julia}
Julia: {https://julialang.org/}

\bibitem{aligator}
Kov{\'{a}}cs, L.: {Aligator: {A} Mathematica Package for Invariant Generation
  (System Description)}. In: {IJCAR}. Lecture Notes in Computer Science,
  vol.~5195, pp. 275--282. Springer (2008)

\bibitem{sympy}
Meurer, A., Smith, C.P., Paprocki, M., \v{C}ert\'{i}k, O., Kirpichev, S.B.,
  Rocklin, M., Kumar, A., Ivanov, S., Moore, J.K., Singh, S., Rathnayake, T.,
  Vig, S., Granger, B.E., Muller, R.P., Bonazzi, F., Gupta, H., Vats, S.,
  Johansson, F., Pedregosa, F., Curry, M.J., Terrel, A.R., Rou\v{c}ka, v.,
  Saboo, A., Fernando, I., Kulal, S., Cimrman, R., Scopatz, A.: {SymPy:
  symbolic computing in Python}. PeerJ Computer Science  \textbf{3}, ~e103
  (2017). \doi{10.7717/peerj-cs.103}

\bibitem{kapur}
Rodr{\'{\i}}guez{-}Carbonell, E., Kapur, D.: {Generating all polynomial
  invariants in simple loops}. Journal of Symbolic Computation  \textbf{42}(4),
   443--476 (2007). \doi{10.1016/j.jsc.2007.01.002}

\bibitem{Sage}
SageMath: {http://www.sagemath.org/}

\bibitem{Mathematica}
Wolfram, S.: {An Elementary Introduction to the Wolfram Language}. Wolfram
  Media Inc. (2017),
  \url{https://www.wolfram.com/language/elementary-introduction/2nd-ed/}

\end{thebibliography}

\end{document}